\documentclass[runningheads]{llncs}
\usepackage{graphicx}
\usepackage{hyperref}
\usepackage{amsmath,amsfonts,amssymb}
\usepackage[dvipsnames]{xcolor}
\usepackage{multirow}
\usepackage{subfig}

\begin{document}

%
\title{Rethink Training of BERT Rerankers \\ in Multi-Stage Retrieval Pipeline}
%
%
\author{
Luyu Gao \and
Zhuyun Dai \and
Jamie Callan
}
\authorrunning{L. Gao et al.}
%
\institute{
Language Technologies Institute, Carnegie Mellon University
\newline
\email{\{luyug, zhuyund, callan\}@cs.cmu.edu}
}
\maketitle              
\begin{abstract}
Pre-trained deep language models~(LM) have advanced the state-of-the-art of text retrieval. Rerankers fine-tuned from deep LM estimates candidate relevance based on rich contextualized matching signals. Meanwhile, deep LMs can also be leveraged to improve search index, building retrievers with better recall. One would expect a straightforward combination of both in a pipeline to have additive performance gain. In this paper, we discover otherwise and that popular reranker cannot fully exploit the improved retrieval result. We, therefore, propose a Localized Contrastive Estimation (LCE) for training rerankers and demonstrate it significantly improves deep two-stage models.\footnote{Our codes are open sourced at \url{https://github.com/luyug/Reranker}.}
\end{abstract}
\section{Introduction}

Recent state-of-the-art retrieval systems are pipelined, consisting of a first-stage heuristic retriever such as BM25 that efficiently produces an initial set of candidate results followed by one or more heavy rerankers that rerank the most promising candidates~\cite{Nogueira2019MultiStageDR}.  Neural language models~(LM) such as BERT~\cite{BERT} have had a major impact on this architecture by providing more effective index terms~\cite{doc2qry} and term weights~\cite{HDCT} for heuristic retriever and providing rich contextualized matching signals between query and document for rerankers~\cite{Dai2019Deeper,Nogueira2019PassageRW}.  


Intuitively, a better initial ranking provides later stage neural rerankers with more relevant documents to pull up to the top of the final ranking. In a perfect world, a neural reranker recognizes the relevant documents in its candidate pool, inheriting all of the successes of previous retriever. However, simply forming the pipeline by appending a BERT reranker to an effective first-stage retriever does not guarantee an effective final ranking. An improved candidate list sometimes causes inferior reranking. 
When the candidate list improves, false positives can become harder to recognize as they tend to share confounding characteristics with the true positives. A discriminative reranker should be able to handle the top portion of retriever results and avoid relying on those confounding features. 

In this paper, we introduce Localized Contrastive Estimation (LCE) learning. We \emph{localize} negative sample distribution by sampling from the target retriever top results. Meanwhile, we use a \emph{contrastive} form loss which penalizes signals generated from confounding characteristics, preventing the reranker from collapsing.  


Experiments on the MSMARCO document ranking dataset show that LCE can better exploit the LMs capability. With the same BERT model, LCE achieves significantly higher accuracy without incurring training or inference overhead. 


\section{Background}
Separation of retrieval into stages was introduced naturally due to efficiency-effectiveness trade-off among different ranking models: fast but less accurate model~(e.g. BM25) retrieves from the entire corpus while slower but more accurate ones~(e.g. BERT) refines ranking in the top candidate list. 

Heuristic retrievers like BM25 use matching signals exclusively from exact match and therefore can use inverted list data structure for low latency full corpus retrieval. They are limited by document statistics for scoring. As a fix, deep language models can be leveraged to re-estimate term weights in search index~\cite{DeepCT,HDCT}. 
An alternative is adding probable query terms to document~\cite{doc2qry}. 

Pre-trained deep LMs~\cite{ELMo,BERT}  have demonstrated strong supervised transfer performance on reranking tasks. Popular recent works~\cite{Dai2019Deeper,Nogueira2019PassageRW} fine-tune BERT~\cite{BERT} with binary classification objective and show it significantly outperforms earlier models. In this paper, we however question if this simple paradigm is sufficient to realize BERT's full potential, especially for high performance deep retrievers that generate candidates consisting of harder negatives.

Alternatives to binary classification objective are the contrastive learning objectives that directly take negatives into account~\cite{Hadsell2006DimensionalityRB}. The popular NCE loss computes scores of a positive instance and several negatives instances, normalize them into probabilities and train the model to give higher probability to the positive instance~\cite{Wu2018UnsupervisedFL}. The incorporation of negatives in loss prevents the model from collapsing. While contrastive loss has been widely studied in representation learning~\cite{Wu2018UnsupervisedFL,Chen2020ASF}, there are few prior works adopting it to train deep LM rerankers.

\section{Methodologies}
\label{sec:method}

\paragraph{Preliminaries}
We aim to train a BERT reranker to score a query document pair,
\begin{equation}
    s = \text{score}(q, d) = \mathbf{v}_\text{p}^\intercal \; cls(\texttt{BERT}(\textbf{concat}(q,d)))
    \label{eq:score}
\end{equation}
where $cls$ extracts BERT's [CLS] vector and $\mathbf{v}_\text{p}$ is a projection vector. 
We refer to the training technique popularly adopted~\cite{Dai2019Deeper,Nogueira2019PassageRW} as the \emph{Vanilla method}. 
It samples query document pairs independently and compute on each individual query-document pair using binary cross entropy~(BCE) based on query $q$ document $d$ and corresponding label~(+/-),
\begin{equation}
    \mathcal{L}_v := \begin{cases}
    \text{BCE}(\text{score}(q, d), +) & \text{d is positive} \\
    \text{BCE}(\text{score}(q, d), -) & \text{d is negative}
    \end{cases}
    \label{eq:balance}
\end{equation}
Vanilla method treats reranker training as a general binary classification problem. However, reranker is unique in nature; 
it deals with the very top portion of retriever results, each of which may contain many confounding signatures. The reranker is expected to,
\begin{itemize}
    \item Exile at handling top portion of retriever results.
    \item Avoid collapsing onto matching with confounding features.
\end{itemize}
To this end, in this section, we introduce Localized Contrastive Estimation~(LCE) loss. The contrastive loss prevents collapsing and localized negative samples focus the reranker on top retriever results.

\paragraph{Localized Negatives from Target Retriever}  
Given a target initial stage retriever and a set of training queries, we use the retriever to retrieve from the entire corpus, generating a set of document rankings for the queries. For each query $q$ then sample from the set $R_q^m$ of top ranked $m$ documents, $n$ non-relevant documents as negatives examples. All sampled documents together form the negative training set. As will be shown in \autoref{sec:analysis}, re-building training set based on the specific target retriever 
is critical to ensure robust training. 


\paragraph{Contrastive Loss} 
After aggregating all negatives sampled from target retriever, we form for each query $q$ a group $G_q$ with a single relevant positive $d^+_q$ and sampled non-relevant negative documents from $R_q^m$. We treat the BERT scoring function as a deep distance function,
\begin{equation}
    dist(q, d) = \text{score}(q, d) = \mathbf{v}_\text{p}^\intercal \; cls(\texttt{BERT}(\textbf{concat}(q,d)))
\end{equation}
with which we define the contrastive loss for one query $q$ as,
\begin{equation}
    \mathcal{L}_q
    := -log \; \frac{exp(dist(q,d^+_q))}{\sum_{d \in G_q} exp(dist(q,d))}
\end{equation}
Importantly, here loss and gradient condition not only on the relevant pair but also the retrieved negatives. This effectively helps prevent collapsing onto simple confounding matchings.

\paragraph{LCE Batch Update}
Putting it all together, we can define the Localized Contrastive Estimation (LCE) loss on a training batch of a set of query $Q$ as,
\begin{equation}
    \mathcal{L}_\text{LCE}
    := \frac{1}{|Q|} \sum_{q \in Q, G_q \sim R^m_q} -log \; \frac{exp(dist(q,d^+_q))}{\sum_{d \in G_q} exp(dist(q,d))}
\end{equation}
Compared to a standard noise contrastive estimation~(NCE) loss, LCE uses the target retriever to localize negative samples and focus learning on top portion instead of randomly sampled noisy negatives.

\section{Experiment Methodologies}
\textbf{Dataset and Tasks}
We use the MSMARCO~\cite{MSMARCO} \emph{document} ranking dataset. The dataset contains 3 million documents. A document consists of 3 fields (title, URL, and body) with around 900 words. Models are trained on the train set of 0.37M training pairs. As recommended by MSMARCO organizers, we use the dev set for analysis. 

\paragraph{Initial Stage Retriever}
We experimented with four initial retrievers: Indri, un-tuned BM25, tuned BM25 (denoted as BM25*), and HDCT~\cite{HDCT}. The Indri search results come from MSMARCO organizers\footnote{\url{https://microsoft.github.io/msmarco/}}. We build BM25 indices with the Anserini toolkit~\cite{Anserini}, from which we produce two sets of search results with the toolkit's default BM25 parameters and a set of tuned parameters suggested by the toolkit authors\footnote{\url{https://github.com/castorini/anserini}}. 
HDCT is the SOTA method for augmenting document search indices with term weights re-estimated with BERT; we use the rankings provided by the authors \footnote{\url{http://boston.lti.cs.cmu.edu/appendices/TheWebConf2020-Zhuyun-Dai/}}. 
We input top 100 candidate lists to rerankers. 
\paragraph{Implementation}
Following \cite{Dai2019Deeper}'s BERT-FirstP setup, we input the concatenated document title, url and body's first 512 queries to the rerankers. Our rerankers are built and trained in mixed precision with PyTorch~\cite{pytorch} and based on Huggingface's BERT implementation~\cite{hf-transformers}. We sample negatives from the target retriever's top ranked $m=100$ documents similar to reranking depth. We train on 4 RTX 2080 ti GPUs, each with a batch of 8 documents. We train for 2 epochs, with a 1e-5 learning rate and a warmup portion 0.1.

\section{Document Ranking Performance}
\label{sec:perf}

\begin{table}[t]
  \centering
  \caption{
  Document ranking performance measured on MSMARCO dev (left table)
  and eval set (right table). $\dagger$ indicates statistical significance over Vanilla using a t-test with $p<0.05$. As the leaderboard eval set only reports aggregated metrics, we cannot report statistical significance.
  }
  \subfloat{%
    \begin{tabular}{ l |llll }
    \hline \hline 
    \multirow{2}{*}{Method} & \multicolumn{4}{c}{MSMARCO Dev}  \\ 
     & \multicolumn{4}{c}{MRR@100}  \\  
     \hline
     & Indri & BM25 & BM25* & HDCT \\
     \hline
     Vanilla & 38.34 & 36.97 & 39.28 & 40.84 \\
     LCE & 39.55$^\dagger$  & 39.66$^\dagger$  & 42.23$^\dagger$  & 43.38$^\dagger$  \\
     \hline \hline 
    \end{tabular}
  }
  \subfloat{%
     \hspace{.5cm}%
    \begin{tabular}{ l | c }
    \hline \hline 
    \multirow{2}{*}{Method} & MSMARCO Eval  \\ 
     & MRR@100  \\  
     \hline
     PROP (ensemble) \footnotemark & 40.1 \\
     BERT-m1 (ensemble) \footnotemark & 39.8 \\
     \hline
     Indri + Vanilla & 33.8 \\
     HDCT + LCE (single) & 38.2  \\
     HDCT + LCE (ensemble) & \textbf{40.5} (1st place) \\

     \hline \hline 
    \end{tabular}
`  }
  \label{tab:ranking}
  \vspace{-4mm}
\end{table}

In table~\ref{tab:ranking}, we summarize ranking performance on MSMARCO document ranking Dev and Eval~(leaderboard) queries. Here, both vanilla and LCE use negatives from target retriever.
On the dev set, we test rerankers trained with vanilla and LCE loss on each type of the first-stage retriever. We see LCE significantly improves performance with all retrievers. Meanwhile, we see that gain using LCE enlarges as the retriever grows stronger, suggesting it can capture more complicated matching in the improved candidate list, while not being confused by the harder negatives. 


\addtocounter{footnote}{-2} 
\stepcounter{footnote}\footnotetext{PROP\_step400K base (ensemble v0.1)}
\stepcounter{footnote}\footnotetext{BERT-m1 base + classic IR + doc2query (ensemble)}

The leader board results confirmed the effectiveness of LCE. HDCT+LCE pipeline outperformed the vanilla basline by a large margin.
Following other recent leaderboard submissions, we further incorporate model ensemble. Our ensemble entry uses an LCE trained ensemble of BERT, RoBERTa~\cite{Liu2019RoBERTaAR} and ELECTRA~\cite{Clark2020ELECTRAPT} to rerank HDCT top 100. This submission got first place, achieving the state-of-the-art performance\footnote{On the camera ready date (January 20th, 2021).}.

\section{Analysis}
In this section, we first analyze the effect of number of sampled documents per query in LCE, then the influence of the negative sample localization. 
\label{sec:analysis}
\begin{figure} [h]
    \vspace{-6mm}
    \centering
    \includegraphics[width=.99 \textwidth]{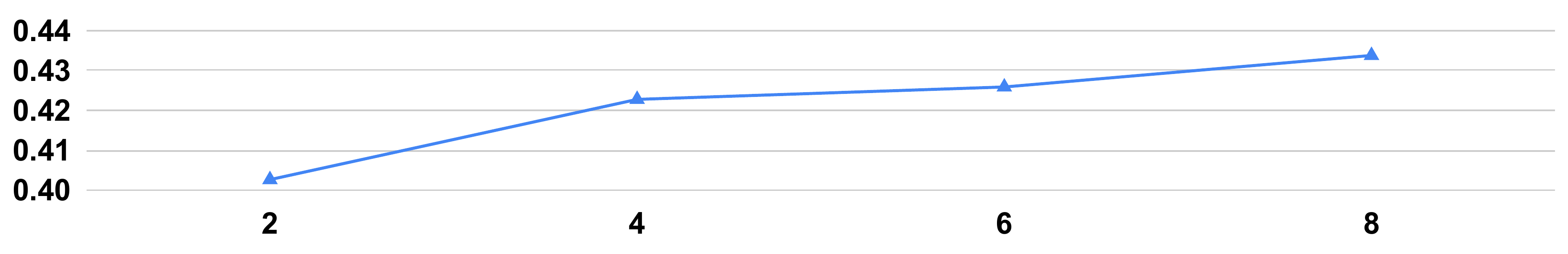}
    \caption{Effect of LCE sample size We plot MRR@100 against sizes.}
    \label{fig:group-size}
\end{figure}
\vspace{-8mm}
\begin{figure}[t]
    \includegraphics[width=.99 \textwidth]{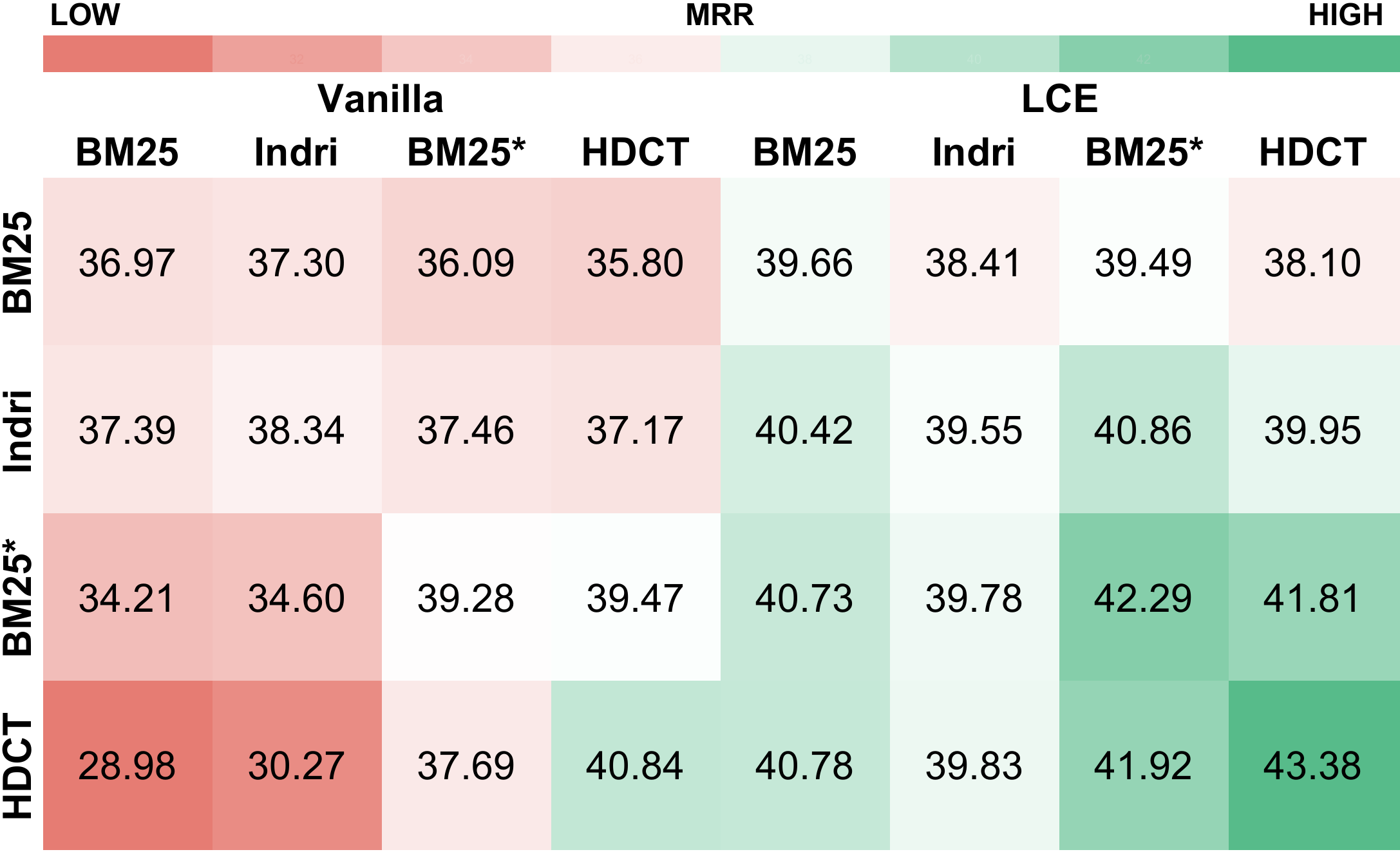}
    \caption{Effects of Train Retriever. The \textbf{horizontal} axis is the retriever that generates negatives for training (train retriever); the \textbf{vertical} axis is the retriever that generates candidates for testing (test retriever).
    }
    \label{fig:paring}
\end{figure}
\paragraph{Effect of LCE Sample Size} In \autoref{fig:group-size}, we study the effect of varying the number of sampled documents per query in the LCE loss. We observe a big improvement from size 2~(1 positive, 1 negative) that compute loss scale with a single negative, to size 4 where loss weights are computed with 3 negatives. Further increase in sample size can generate some additional improvements. 


\paragraph{Influence of Negative Localization} LCE samples negatives from top ranked documents retrieved by target retriever. Here we quantitatively evaluate its importance. Denote retriever used in training for negative sampling \emph{train retriever} and in testing for candidate generation \emph{test retriever}. We use all rerankers from \autoref{sec:perf} to rank candidate lists generated by all retrievers and plot results in a heat map~\autoref{fig:paring}. We plot rerankers using different train retrievers on the \textbf{horizontal} axis and test retrievers on the \textbf{vertical} axis. Each $4\times4$ sub-grid corresponds to a training strategy, and sub-grid diagonals correspond to \autoref{sec:perf} results. 
We observe localization benefits both LCE and vanilla methods. 
The performance of the Vanilla trained reranker \emph{drops} severely when negatives are not localized by test retriever but from a weaker train retriever. Similarly, rerankers trained with LCE loss also perform better with localization. 
Interestingly, we do find that the LCE loss can bring some degrees of adaptability to the reranker, making it robust when the test retriever is different from the train retriever.
\section{Conclusion}




Recent research shows promising results on using deep LMs to improve initial retrievers. However, we discovered that previous BERT rerankers could not fully exploit the improved initial rankings.
We propose Localized Contrastive Estimation~(LCE) learning, to localize training negatives with target retriever, 
and to use a contrastive loss to penalize matching with confounding characteristics.

Experimental results demonstrate that reranker trained with LCE significantly outperforms its vanilla method trained counterpart using the same LM. 
Our analysis shows that localizing negatives and having an expressive loss with multiple contrastive negatives are both critical for the success of LCE.

The positive results show that, instead of adopting more advanced LM, it is also possible to improve the performance of existing deep LMs with better learning methods.
Meanwhile, before this work, there are few existing work studying the interaction between different deep retrievers and reranker in pipelined retrieval systems. We believe this paper will encourage the community to conduct more systematic research on pipelined IR systems. 

%
%
%
\newpage
\bibliography{main}
%




\end{document}